\documentclass{ptptex}
\usepackage{graphicx}
\usepackage{amssymb}
\usepackage{wrapft}
\def\half{{\textstyle\frac{1}{2}}}
\newcommand{\bR}{{\mathbb R}}
\def\e{\mathop{\mathrm{e}}\nolimits}
 \def\Cset{\mathbb{C}}
\newcommand{\be}{\begin{equation}}
\newcommand{\ee}{\end{equation}}
\newcommand{\hh}{\mathcal J}
\newcommand{\n}{{\bf n}}
\newcommand{\m}{{\bf m}}
\newcommand{\rr}{{\bf r}}
\newcommand{\hs}{{\hat \sigma}}
\newcommand{\ii}{{\mathrm{i}}}
\def\sc#1#2{\langle #1 , #2 \rangle }

\def\j{{\mathcal J}(V)}
\def\vp{V_{+}}
\def\lp{{\mathcal L}_{+}(V)}
\def\ss{{\mathcal S}}
\def\aa{{\mathcal A}}
\def\hh{{\mathcal H}}
\def\pr{{\mathcal P}_{+}}
\newcommand{\aaa}{A_\alpha}
\newcommand{\gba}{g_{\beta\alpha}}

\newcommand{\la}{\Lambda_\alpha}
\newcommand{\gab}{g_{\alpha\beta}}
\newtheorem{thm}{Theorem}
\newtheorem{lem}[thm]{Lemma}
\newtheorem{defn}[thm]{Definition}
\newtheorem{algorithm}{PDP Algorithm}
\DeclareGraphicsExtensions{.jpg,.pdf,.gif,.png,.eps}
\begin{document}
\markboth{Arkadiusz Jadczyk}{Quantum Fractals on the Poincar\'e Disk}
\title{Piecewise Deterministic Quantum Dynamics and Quantum Fractals on the Poincar\'e Disk.}
\author{A. Jadczyk }
\inst{ Institute of Theoretical Physics, University of Wroclaw,\\ Pl. Maxa Borna 9, 50 204 Wroc{\l}aw, Poland\footnote{On leave of absence. Email: ajad@th-phys.edu.hk}}
\maketitle
\abst{
It is shown that piecewise deterministic dissipative quantum dynamics in a vector space with indefinite metric can lead to well defined, positive probabilities. The case of quantum jumps on the Poincar'e disk is studied in details, including results of numerical simulations of quantum fractals.}
\section{Introduction}
\begin{wrapfigure}{r}{6.6cm}
  
      \includegraphics[width=6.6cm, keepaspectratio=true]{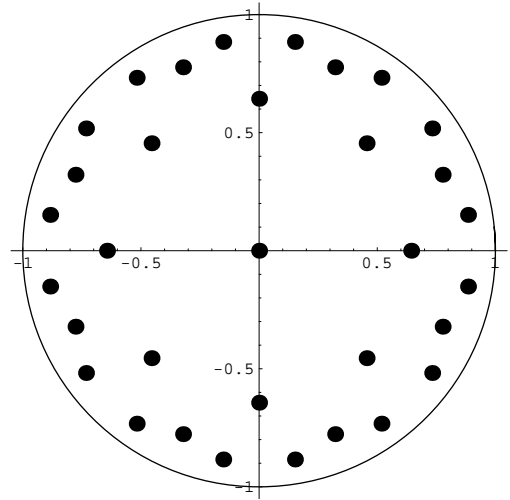}
\caption{33 points from the hyperbolic tiling of the Poincar\'e disk with Schlafli symbol $(8,3)$}\label{fig:circle}
\end{wrapfigure}

\begin{figure}[!htb]
  \begin{center}
    \leavevmode
      \includegraphics[width=9cm, keepaspectratio=true]{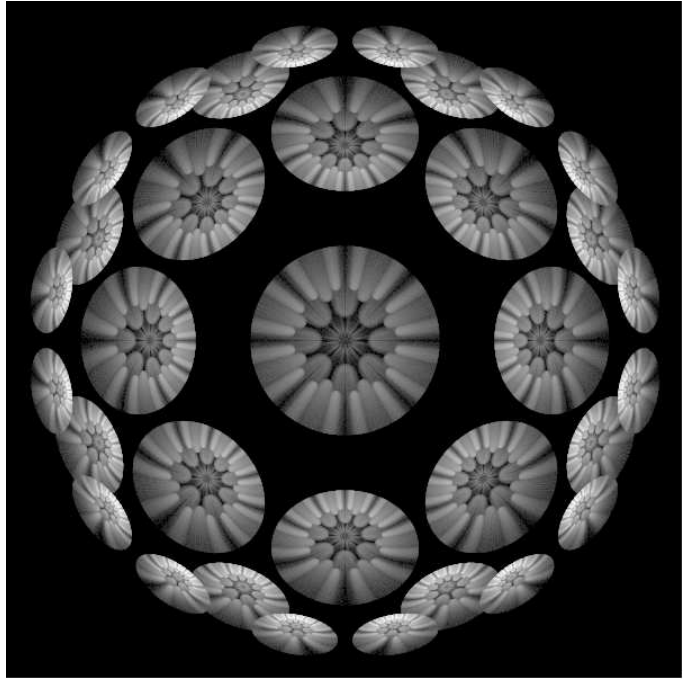}
  \end{center}
\caption{Hyperbolic quantum fractal $\epsilon=0.75.$}\label{fig:hyper075}
\end{figure}

\begin{figure}[!htb]
  \begin{center}
    \leavevmode
      \includegraphics[width=9cm, keepaspectratio=true]{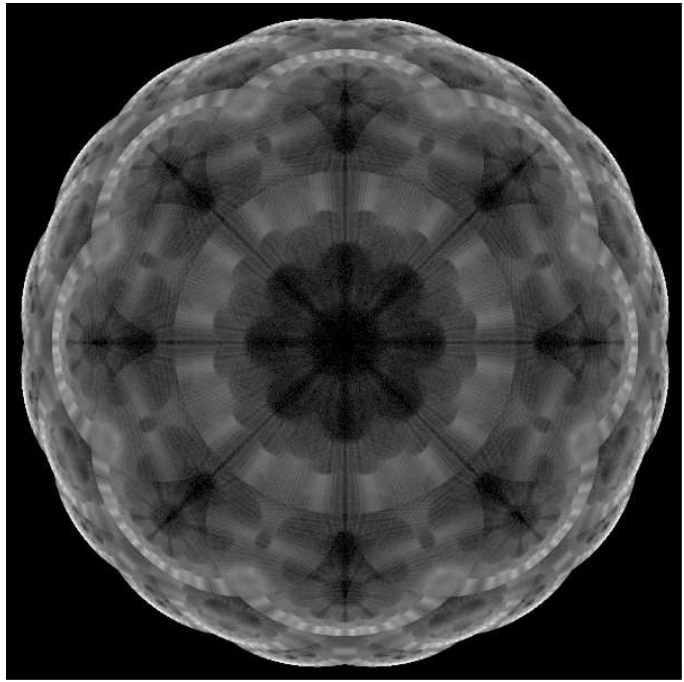}
  \end{center}
\caption{Hyperbolic quantum fractal $\epsilon=1.75.$}\label{fig:hyper175}
\end{figure}

One of the most important characteristic features of a quantum theory, as opposite to a pure classical theory, is that the set of pure states of a quantum system is not a simplex. Consider a quantum system described in a complex Hilbert space $\hh.$ Its pure states are described rays $\{e^{\ii\phi}\psi :\phi\in[0,2\pi]\}\subset\hh,\, ||\psi||=1$ or, equivalently, by one-dimensional 
orthogonal projections $P=|\psi><\psi |.$ Let $\ss$ be the set of all pure states.  If $\mu$ is a probabilistic measure on $\ss$ then, according to the standard, linear, quantum mechanics, only the density matrix $\rho=\int_\ss Pd\mu (P)$ is observable, not the measure $\mu$ itself. And yet, as demonstrated in Ref. \citen{jad94b}, while searching for the dynamical mechanism of particle track creation and event generation, the measurement process introduces a mild non-linearity and leads to a unique random process on $\ss$ - provided the quantum system is coupled, in an appropriate way, to a classical system. \footnote{A similar approach has recently been discussed by A. Peres \cite{peres1,peres2}} This mild non-linearity is related to the ``time-of-events" observable studied in Ref. \citen{blaja95f}, and calls for a careful re-examination of the fundamental axioms of quantum mechanism. In particular, if the Event Enhanced Quantum Theory (EEQT) developed by Blanchard an the present author will prove to be correct, then precise measurements of timing of events may give us more information about the actual state of the quantum system than it is usually assumed. The present paper, although stemming from the above ideas, deals with a pure formal aspect of EEQT - namely with a piecewise deterministic process of jumps (state reductions) on the space of pure states in a space with indefinite metric. Spaces with indefinite metric are usually being considered as too pathological for applications in quantum theory \cite{araki85}. And yet within the formalism of EEQT, where none of the standard probabilistic axioms of quantum theory are required, and that because all the interpretation comes from a dynamical coupling between a classical and a quantum systems, it is possible to have a well defined dynamics with all positive probabilities. In the present paper we will follow some of the ideas and notations developed in Ref. \citen{jad71}, though we will make an effort at presenting this paper as self-contained. We will start with a linear Linblad's type master equation for observables in a Krein space, with a particular kind of coupling between a quantum system and a finite-state classical system -- cf. Eq. (\ref{eq:lioua}). We skip the possible unitary Hamiltonian part of the evolution, and we concentrate on the dissipative part alone. By Theorem \ref{thm:2} the coupling generates a unique, piecewise deterministic process on the space of pure states of the total system. Next, we farther specify the coupling in such a way that we can take the trace over the classical system and get the effective piecewise deterministic process on the hyperboloid of pure states (of positive norm squared) of the quantum system - cf. PDP Algorithm in Section \ref{sec:lin}. We then examine in details the particular case of Krein's space $V_{1,1}$ when the space of pure states is isomorphic to the Poincar\'e
disk. We specify the coupling operators to be fuzzy projections (known also as ``spin coherent states") as defined in Eq. (\ref{eq:fuzzy}), on a family of symmetrically distributed states on the disk (a selection of 33 points from the hyperbolic tiling with Schlafli symbol (3,8) - cf. Fig. \ref{fig:circle}). The plot of the sequence of 100 mln of quantum jumps generated by the PDP process suggests that there is a fractal attractor set. Figures \ref{fig:hyper075} and \ref{fig:hyper175} show the results if the iteration process for two different values of the fuzziness parameter $\epsilon=0.75,\epsilon=1.75$ (for $\epsilon=1.0$ the projections are sharp). Due to the fact that our process is a hybrid one - it has jumps with place-dependent probabilities, as in iterated function systems, but it also has dissipative parts of a continues evolution with a random time length, and also due to the fact that the Poincar\'e disk is non-compact, there are no ready to use theorems that we could apply in order to ascertain the existence and uniqueness of the invariant measure, as it was done in the case of positive definite metric\cite{jad03b}. Our numerical simulations suggest a conjecture that the probabilities of getting outside a bounded region decrease fast enough with the dimension of this region, so that the 
attractor set and the invariant measure supported by this set are well defined. It should be mentioned that, according to the linear paradigm of quantum mechanics, mentioned at the beginning of this section, the patterns displayed in the figures, although mathematically unique, are physically unobservable. According to this paradigm, the only observable is the statistical matrix resulting from the integration of projections with respect to the measure. In our case, due to symmetry of the pattern, it is clear that this statistical state is a mixture of the maximal entropy state $\frac{1}{2}I,$  and the orthogonal projection on the eigenstate of $\sigma_3$ corresponding to the eigenvalue $1.$ It is worthwhile to notice that Lozinski et al.\cite{loslzy} recently studied iterated function systems on the space of statistical operators of a given quantum system. Their approach, even though different in spirit, has several areas of overlap with piecewise deterministic processes generated by EEQT.

\section{Indefinite metric space}
Let $V$ be a complex vector space of finite dimension $n$ with a
non degenerate sesquilinear form $\sc{\psi }{\phi}.$\footnote{It should be noted that the term ``Hilbert space" is usually reserved for a space with a positive definite scalar product. The term ``Krein space" though less known, is more appropriate here.} Let $L(V)$ be
the algebra of all linear operators on $V$ endowed with the star
operation defined by $\sc{A\psi}{\phi}=\sc{\psi }{A^\star v}.$ $A\in
L(V)$ is said to be {\em symmetric\/ } if $A=A^\star .$ $U\in
L(V)$ is said to be unitary if $U^\star = U^{-1}.$ Let ${\mathcal
H}(V)$ and ${\mathcal U}(V)$ denote the set of all symmetric and
all unitary operators resp. Let ${\mathcal J}(V)$ be the set of
all $J\in {\mathcal H}(V)\bigcap {\mathcal U}(V)$ such that
the sesquilinear form $(\psi,\phi )_J\doteq\sc{\psi }{J\phi}$ is
positive definite. We will assume that ${\mathcal J}(V)$ has at
least two elements. This is equivalent to assuming that the scalar
product $\sc{\psi }{\phi}$ is indeed indefinite, that is that
there exists a basis $\{e_1,\ldots ,e_k,e_{k+1},\ldots ,e_{k+l}\}$
such that $\sc{e_\alpha}{e_\beta}=g_{\alpha \beta}$ where
\be
\{g_{\alpha \beta}\}=\mathrm{diag}(+1,\ldots ,+1,-1,\ldots ,-1)
\ee
with $k,l\geq 1 .$ Whenever we will need to specify the signature of $V,$ we will use the notation $V_{(k,l)}$ for $V.$ There is one-to-one correspondence between
elements of ${\mathcal J}(V)$ and $k$-dimensional subspaces of $V$
on which $\sc{\psi }{\phi}$ is positive definite. The group
${\mathcal U}(V)$ is isomorphic to $U(k,l)$ and acts transitively
on $\j$ with the fixpoint group isomorphic to $U(k)\times U(l),$
thus
\be
\j\equiv \frac{U(k,l)}{U(k)\times U(l)}=\frac{SU(k,l)}{S\left( U(k)\times U(l)\right)}.
\ee
In particular, for $k=l=2$ we get the Cartan symmetric domain $A(2)\equiv D_4\equiv SO(4,2)/S\left( O(4)\times O(2)\right),$ with Shilov boundary isomorphic to the compactified Minkowski space, while for $k=l=1$ we get 
the Poincar\'e disk $A(1)\equiv D_1\equiv SO(1,2)/SO(2)$ whose Shilov boundary is a circle, with a physical interpretation of it being a compactified, circular, time.

Given $J\in\j$ we have positive-definite scalar product $(\psi
,\phi )_J\doteq \sc{\psi }{J\phi }.$ Let us denote by $A^J$ the
Hermitian conjugation with respect to this scalar product. We call
$A\in {\mathcal L}(V)$ $J-$Hermitian if $A=A^J.$ It is then easy
to see that $A\mapsto JA$ gives a one-to-one correspondence
between symmetric and $J-$Hermitian operators. 
\begin{defn}
We will use the following notation:
\begin{eqnarray}
\vp&=&\{\psi\in V: \sc{\psi}{\psi}\geq0,\\
\lp&=&\{A\in {\mathcal L}(V): AV_{+}\subset V_{+}\}
\end{eqnarray}
For $\psi\in\vp$ we will write $\Vert\psi\Vert$ for $\sqrt{\sc{\psi}{\psi}}.$
\end{defn}
Notice that $A\in\lp$ if and only if $\Vert A\psi\Vert^2\geq 0$ for all $\psi\in\vp.$
 Each $J\in\j$ is
a difference of two orthogonal, complementary projections:
\be
J=E-F=2E-I,
\label{eq:jemf}
\ee
where $E=E^\star=E^2$ projects on a $k$-dimensional subspace on
which $\sc{\psi }{\phi}$ is positive definite, and $F=F^\star=I-E$
projects on the complementary $l$-dimensional subspace, where
$\sc{\psi }{\phi}$ is negative definite.
\section{Linblad's type semigroups and associated piecewise deterministic processes in $V_{(k,l)}.$\label{sec:lin}}
It is well known that certain types of quantum dynamical semigroups \cite{alicki87,alicki02} are in one-to-one 
correspondence with piecewise deterministic Markov processes on the space of pure states of the algebra of observables of the system - cf. \cite{blaja95a,blaol99} and references therein). By analyzing the proof of the existence and uniqueness theorems it is easy to see that these results extend to $V_{k,l}$ provided the operators that implement the jumps are in $\lp.$
\begin{thm}
Let $\ss$ be a finite set. Let $V=V_{k,l},$ and let $\aa$ be the $\star$-algebra $\aa=\bigoplus_{\alpha\in\ss} {\mathcal L}(V).$ A typical element of $\aa$ is a sequence $A=(A_\alpha )_{\alpha\in\ss},\ A_\alpha\in {\mathcal L}(V).$ For each pair $\alpha,\beta\in\ss,\ \alpha\neq\beta ,$ let there be given an operator $g_{\alpha\beta}\in\lp .$
Consider semigroup of linear maps $\phi_t(A)_\alpha\doteq\exp (Lt):\aa\rightarrow \aa,t\geq 0,$ with infinitesimal generator $L$ defined by

\be 
L(A)_\alpha=\sum_{\beta\neq\alpha} \gba^\star A_\beta\ \gba - \frac{1}{2}(\la\aaa+\aaa\la ),\label{eq:lioua}
\ee
where
\be
\la=\sum_{\beta\neq\alpha} \gba^\star \gba.\ee
The formula
\be
\mathrm{tr}\left(P\phi_t(A)_\alpha\right)=\sum_\beta \int \mathrm{tr}(AQ)dp(t;P,\alpha;Q,\beta )
\ee
gives then one-to-one correspondence between semigroups $\phi_t$ and piecewise deterministic Markov processes
on $\pr\times\ss$ with transition probabilities $p(t;\alpha,P;\beta,dQ)$ described by the following algorithm:\\
Suppose that at time $t_0$ the system is described by a pair $(P\in\pr,\alpha\in\ss),$ with $P$ bring an orthogonal projection on a vector $\psi\in\vp.$ Choose
a uniform random number $p\in [0,\,1]$, and proceed with the continuous
time evolution by solving the differential equation
\be \dot{\psi}_t\:=\:\exp(-\:\frac{1}{2}\la )\:\psi_t\label{eq:psit}\ee
with the initial vector $\psi$ until $t\,=\,t_1$, where $t_1$ is
determined by
\be \Vert \psi_t\Vert^2\:=\:p\label{eq:int}\ee
Then jump. When jumping, change $\alpha\to\beta$ with probability
\be p_{\alpha\to\beta}\:=\:\|\gba\psi_{t_1}\|^2/(\psi_{t_1},\,\la\psi_{t_1})\label{eq:p}\ee
and, if $p_{\alpha\to\beta}\neq 0,$ change
$$\psi_{t_1}\to\psi_1\:=\:\gba\psi_{t_1}/\|\gba\psi_{t_1}\|.$$
Repeat the steps substituting $t_1,\,\psi_1,\beta$ for $t_0,\,\psi_0,\,\alpha .$ 
$\beta$.\label{thm:2}
\end{thm} 
{\bf Proof}
Algebraically, the proof goes, step by step, exactly the same way as in \cite{blaja95a,blaol99} (cf. also \cite{jakol95} for the uniqueness), the only thing that we need to check is that positivity conditions required in the Hilbert space case also hold when the scalar product is indefinite. If $f(t)\doteq\Vert\psi_t\Vert^2,$ then the function $f(t)$ is real analytic, with $f(0)=1$ and ${\dot f}(t)=-\sc{\psi_t}{\la\psi_t}.$ As long as $f(t)>0$, that is as long as $\psi_t\in\vp$, the derivative ${\dot f}(t)$ is non-positive, because $\sc{\psi_t}{\la\psi_t}=\sum_{\beta\neq\alpha}\Vert \gab\psi_t\Vert^2,$ and $\gab\in\lp.$ Therefore $f$ is monotonically decreasing from $f(0)=1$ and either it never reaches the given value $p\in(0,1),$ so the jump never happens, or, if it reaches the value $p$ at a finite time $t_1,$ then $\psi_{t_1}\in\vp$ and therefore, because $\gab\in\lp$, the jump probabilities $p_{\alpha\to\beta}$ are non-negative. Finally, notice that whenever the denominator in Eq. (\ref{eq:p}) vanishes, this can happen only for discrete values of $t_1,$ thus on a set of measure zero.$\Box$\\

Here, as in Ref. \citen{jad03b} we will be interested in a special case of the above process that leads to simple iterated function systems with place-dependent probabilities, as discussed in \cite{barnsley}. Let $N$ be a natural number, and let $\ss=2^N$ be the set of all binary sequences of length $N.$ Assume  $\lp\ni\gab=g_i\neq 0$ when $\alpha$ differs from $\beta$ only at one, the $i$-th bit, otherwise $\gab=0.$ Let \be\Lambda=\sum_{i=1}^N g_i^\star g_i.\ee If we take a trace over the index $\alpha,$ we end up with a piecewise deterministic process on $\pr$ that can be described as follows:\\
\begin{algorithm}
Start, at time $t_0$, with a vector $\psi\in\vp.$ Choose
a uniform random number $p\in [0,\,1]$, and proceed with the continuous
time evolution by solving the differential equation
\be \dot{\psi}_t\:=\:\exp(-\:\frac{1}{2}\Lambda )\:\psi_t\label{eq:psitg}\ee
with the initial vector $\psi$ until $t\,=\,t_1$, where $t_1$ is
determined by
\be \Vert \psi_t\Vert^2\:=\:p\label{eq:inti}\ee
Then jump. When jumping, change $$\psi_{t_1}\to\psi_1\:=\:g_i\psi_{t_1}/\|g_i\psi_{t_1}\|$$ with probability 
\be p_i(\psi)\:=\:\|g_i\psi_{t_1}\|^2/\sum_{j=1}^N \|g_j\psi_{t_1}\|^2\\ \label{eq:psir}\ee
Repeat the steps replacing $t_0,\,\psi_0$ with $t_1,\,\psi_1.$
\end{algorithm}

\section{Example: $V_{1,1}$ - the Poincar\'e disk.}
As a simple example consider the case of $k=l=1.$ Let us choose
$J_0\in\j$ and let $e_1,e_2$ be an orthonormal basis in $V$
diagonalizing $J_0$. Thus $\sc{e_1}{e_2}=-\sc{e_2}{e_2}=+1,$ $V$
can be identified with $\Cset^2$ - the set of all column vectors
\be
a=\begin{pmatrix}  a_1 \cr
  a_2\end{pmatrix}
\ee
where $a_1$ and $a_2$ are complex numbers, $J_0$ has now the
matrix form:
\be
J_0=\begin{pmatrix}1,&0\cr 0,&-1\end{pmatrix}
\ee
while the positive definite scalar product $(\psi ,\phi )_{J_0}$ is
nothing but the standard
\be (a,b)_{J_0}={\bar a}_1 b_1+{\bar a}_2 b_2
\ee
where the bar in ${\bar c}$ stands for the complex conjugation of
the complex number $c.$ Let us denote by $A^\dag$ the Hermitian
conjugate matrix of $A.$ Using  our previous notation
$A^\dag=A^{J_0}.$ The most general $J_0-Hermitian$ matrix,
$A=A^\dag$ is of the form
\be
A=x^0I+\sigma(\bf x)
\ee
where
\be
\sigma ({\bf x})= x^1\sigma_1+x^2\sigma_2+x^2\sigma_3
\ee
and $\sigma_i$ are the Pauli matrices:
\be
\sigma_1=\sigma_x=\begin{pmatrix}0,&1\cr 1,&0\end{pmatrix}\ee
\be
\sigma_2=\sigma_y=\begin{pmatrix}0,&-i\cr i,&0\end{pmatrix}\ee
\be
\sigma_3=\sigma_z=\begin{pmatrix}1,&0\cr 0,&-1\end{pmatrix}
\ee
and $x^\mu,\ \mu=0,1,2,3$ are real numbers. Note that now $J_0$ is
represented by $\sigma_3 .$ It follows that the most general
symmetric (with respect to $\sc{\phantom u}{\phantom v}$) operator
on $V$ is of the form

\be
\sigma_3 (x^0+\sigma ({\bf
x}))=x^3-ix^2\sigma_1+ix^1\sigma^2+x^0\sigma_3,
\ee
where $x^\mu$ are real. In other words moving between symmetric and
$J-$Hermitian operators is accomplished by the "Wick rotation" $x^1\mapsto -i x^2,x^2\mapsto i x^1 $ in
the parameters $x^1,x^2 .$
\vskip10pt
\noindent
\subsubsection{Fuzzy projections}
In our toy model state of the system is represented
by a unit vector$\phi\in\vp.$ Proportional vectors describe the same state, therefore it is better
to represent states by one dimensional orthogonal projections $P$ on positive subspaces. The set of all states
will be denoted by $\pr.$ Quantum jumps will be
implemented by operators in $\lp.$
In this paper, as in
Ref.\cite{jad03b}, we will be interested in fuzzy projections of
the form \be Q(J,\epsilon)=\half (I+\epsilon J)\label{eq:fuzzy}\ee which have the
above property. Notice that for $\epsilon=1$ we have $Q(J,1)=\half
(I+(2E_J-I))=E_J$, and thus $Q(J,1)$ is a sharp projector. Notice that in $V_{1,1}$ we have $Q(J,1)\in\pr$ because in this case maximal positive subspaces are one dimensional. We want
to know for which values of $\epsilon$,  $Q(J,\epsilon )$ is in $\lp.$ For this it is enough to calculate trace of
$Q(J,1)Q(J^\prime,\epsilon)Q(J,1),$ and to find the conditions on
$\epsilon$ which guarantee its positivity for all $J\in \j .$ Let us
use the representation using $J_0$ and Pauli matrices as above. 
\begin{defn}
For all $\m=(m_0,m_1,m_2), \n=(n_0,n_1,n_2)\in\bR^3$ set $<\m\cdot\n>=m_0n_0-m_1n_1-m_2n_2.$ Let  $\hh=\{\m=(m_0,m_1,m_2): 
<\m\cdot \m >=1,\, m_0>0\} .$
\end{defn}
\begin{lem}\label{jrepres}
Each $J\in \j $ is uniquely representable in
the form
\be
J=J({\m })={\hat \sigma} ({\m } )=
m_0\sigma_3-im_2\sigma_1+im_1\sigma_2,
\ee
where $\m\in\hh .$
\end{lem}
{\bf Proof} $J$ must be symmetric, therefore $J=x^3-ix^2\sigma_1+ix^1\sigma^2+x^0\sigma_3,$ with $x^\mu$ real. $J^2=I$ implies then that $x^3=0$ and $(x^0)^2-(x^1)^2-(x_2)^2=1.$ Now, $<\psi,J\psi >=(\psi,J_0J\psi )_{J_0},$ therefore, for $<\psi,J\psi >$ to be non-negative for all $\psi ,$    $J_0J=\sigma_3 J =x^0+x^1\sigma_1+x^2\sigma_2$ must be positive in the standard sense. That means $x^0 > 0.$
$\Box$\\
\begin{defn}

We will write $Q(\m ,\epsilon)$ for $Q(J(\m ),\epsilon ),$ i.e.
\be
Q(\m ,\epsilon)=\half (I+\epsilon\hs (\m )),
\ee
or, explicitly, in a matrix form:
\be
Q(\m ,\epsilon)=\half\begin{pmatrix}1+\epsilon m_0&\epsilon (m_1-i m_2)\cr
\epsilon (-m_1+i m_2)&1-\epsilon m_0 \end{pmatrix}
\ee
\end{defn}

Note that $\mathrm{Tr}(Q(\m ,\epsilon))=1.$ 
\begin{thm}
\begin{enumerate}
\item[(i)]For each $\m ,\rr\in\hh ,\, \epsilon\in\bR$ we have
\be
 Q(\m ,\epsilon )Q(\rr )Q(\m ,\epsilon )=\lambda(\epsilon,\m,\rr) Q(\rr^\prime),
\label{eq:lambda1}
\ee
where
\be
\lambda(\epsilon,\m,\rr)=\frac{1+\epsilon^2+2\epsilon <\m\cdot\rr >}{4}
\label{eq:lambda2}
\ee
and
\be
\rr^\prime=\frac{(1-\epsilon^2)\rr+2\epsilon(1+\epsilon <\m\cdot\rr> )\m}{1+\epsilon^2+2\epsilon
<\m\cdot\rr> }\in\hh ,
\ee
\item[(ii)] The fuzzy projections $Q(\m ,\epsilon )$ map $V_{+}$
into $V_{+}$ if and only if $\epsilon\geq 0.$
\end{enumerate}
\end{thm}
{\bf Proof}
(i) Follows by inspection, or by applying the ``Wick rotation" to the formulas in Ref. \cite{jad03b}.\\
(ii)  Let $A=A^\star$ be a linear operator on $V$ which maps $V_{+}$ into $V_{+}$, then we must have $\sc{A\psi
}{A\psi }=\sc{\psi}{A^2\psi}\geq 0$ for all $\psi\in V$ with
$\sc{\psi}{\psi}=1.$ Observe that
$\sc{\psi}{A^2\psi}=\mathrm{Tr}(EA^2E)=\mathrm{Tr}(AEA)$, where
$E$ is the orthogonal projection on $\psi .$ Since
\be
\lambda(\epsilon,\m,\rr)=\mathrm{Tr}(Q(\m ,\epsilon )Q(\rr )Q(\m ,\epsilon
)),
\ee
it is enough to analyze the behavior of the function $\lambda.$
The unitary group of $V$ acts
on $\j $ transitively by its natural action: $J\mapsto UJU^\star$.
As the phase of the unitary operator cancels out, we can restrict
ourselves to the special unitary group, in our case $SU(1,1).$ We
have $2:1$ homomorphism $U\mapsto
\Lambda(U)$ of $SU(1,1)$ onto $SO(2,1)$ given by
\be
UJ(\m )U^\star = J(\Lambda \m ).
\ee

 The function $\lambda$ in Eq.(\ref{eq:lambda2}) is
$SO(2,1)$ invariant. Therefore to analyze its behavior we can
always make an $SO(2,1)$ transformation so that $\rr = (1,0,0),$
then, as a function of $\m\in\hh$ we get $4\lambda (\m,\epsilon )=
1+\epsilon^2+2\epsilon m_0,$ which is bounded from below only when
$\epsilon\geq 0.$ If $\epsilon>0,$ then $\lambda (\m,\epsilon ),$ with $\m\in\hh $, attains minimum equal to
$\frac{(1+\epsilon)^2}{4}$ at $\m=(1,0,0).$ 
$\Box$\\
\subsection{Poincar\'e disk model}
The hyperboloid $\j \approx\pr$ is isomorphic to the Poincar\'e disk $D:$
\be
D=\{ z\in \Cset : \vert z \vert < 1\}
\ee
by the isomorphism:
\be
J({\m })={\hat \sigma} ({\m } )= m_0\sigma_3+i m_1\sigma_2-i
m_2\sigma_1\Leftrightarrow z
\ee
where $m_1+i m_2=2z/(1-\vert z \vert^2) .$

The action of the unitary group ${\mathcal SU}(V)$ on $\j$ is
isomorphic to the action of $SU(1,1)$ on the Poincar\'e disk $D$ via
fractional transformations: $A: z\mapsto
(a_{11}z+a_{12})/(a_{21}z+a_{22}),$ the isomorphism being given by
$A=\sigma_2 U \sigma_2.$
\section{Piecewise deterministic process on $\pr.$}
As an example consider the piecewise deterministic process described by equations (\ref{eq:psit})-(\ref{eq:psir}),
with $g_i=\sqrt{\frac{4}{N}} Q(\m[i] ,\epsilon ),$ $i=1,2,\ldots ,N.$
In \cite{jad03b}, where we were studying quantum jumps on the
complex projective space $PC(1)\approx S^2$, we could
neglect completely the continuous part of evolution of the state
vector, as given by Eq.(\ref{eq:psit}), and that because we could choose symmetrically distributed
configuration of fuzzy projections $Q(\n[i] ,\epsilon)$ in such a
way that $\sum_i Q(\m[i] ,\epsilon)^2=\mathrm{const}\cdot I$ due to
$\sum_i
\m[i] = 0.$ Such a choice is not possible in our case, with sphere
replaced by the positive hyperboloid, because vectors on the
positive hyperboloid can not balance to the zero vector. Given a
sequence $\m[1],\m[2],\ldots,\m[N]$, $\sc{\m[i]}{\m[i]}=1$, we have
\be
\sum_{i=1}^N \m[i]= Nc\, \m_c,
\ee
where $c$ is a constant $c>0$ and $\m_c$, is the centroid vector
of unit length square $\sc{\m_c}{\m_c}=1.$ The EEQT algorithm, as
described, for instance, in \cite{blaja95a}, tells us that when
the quantum jumps are implemented by operators
$g_i$ then time evolution
between jumps is implemented by
\be
\psi (t)= R(t)\psi (0) =
\e^{-\frac{2 t}{N}\sum_{i=1}^N Q(\m[i] ,\epsilon )^2 }\psi
(0).
\ee
Due to the fact that $\hs (\m_c)^2=I$, the exponential is easily
calculated:
\be
R(t)=\e^{-\frac{ t}{2}
(1+\epsilon^2)}[\mathrm{cosh}(\epsilon
ct)-\mathrm{sinh}(\epsilon ct)\hs(\m_c)]
\ee
We will choose our configuration so that $\m_c=(1,0,0),$ thus $\hs
(\m_c )=\sigma_3,$ or, in matrix form:
\be
R(t)=\e^{-\frac{ t}{2}
(1+\epsilon^2)}\begin{pmatrix}\e^{-\epsilon c t}&0\cr
0&\e^{\epsilon c t}\end{pmatrix}.
\ee
It is easy to see that when projection $Q(\rr)$ is the projection
on $\psi (0),$ $Q(\rr)=\vert \psi (0)\rangle\langle \psi (0)\vert
$ then
\be
\Vert \psi (t)\Vert^2 = \mathrm{Tr} (R(2t)Q(\rr )),
\ee
which can be easily calculated form the explicit matrix formulas
above:
\be
p(t)\doteq\Vert \psi (t)\Vert^2 = \exp\left(- t (1+\epsilon^2)\right)\left(\cosh
(2\epsilon ct)-m_0\sinh (2\epsilon ct)\right).
\ee
For $\epsilon>0$ and $m_0\geq 1$ The function $p(t)$ is monotonically decreasing from $p(t_0)=1$ to $p(t_1)=0,$ where
$t_0=0$ and $t_1=\left( \log \left((m_0+1)/(m_0-1)\right)\right)/4\epsilon c.$ In numerical simulations of the PDP process the time of jump is calculated by selecting a uniformly distributed random number $p\in (0,1)$, and then numerically solving the equation $p(t)=p.$ One convenient method of doing it is by introducing variables $y=\exp (4\epsilon ct)$ and $s=(1+\epsilon^2+2\epsilon c)/4\epsilon c,$ and solving numerically the equation $2py^s+y(m_0-1)-(m_0+1)=0$ for $y.$ uniformly distributed random number $p\in (0,1)$, and then numerically solving the equation $p(t)=p.$ One convenient method of doing it is by introducing variables $y=\exp (4\epsilon ct)$ and $s=(1+\epsilon^2+2\epsilon c)/4\epsilon c,$ and solving numerically the equation $2py^s+y(m_0-1)-(m_0+1)=0$ for $y.$ 
To illustrate the resulting iterated function system we took $N=33$ points $\m[i]$ from a hyperbolic tiling with Schlafli symbol $(8,3)$ and plot the results of 100 mln jumps for two $\epsilon=0.75$ and $\epsilon=1.75.$

\newpage
\section*{Acknowledgments} The work would like to thank QFS, Inc. for support.

\end{document}